# Frame Rate, Not the Grid: A Confound in Single-Clip Optical ENF Video Forensics

Ryan Wyton and Bacha Rehman

***Abstract*****: *The optical electric network frequency, the faint mains-driven flicker imprinted on rolling-shutter video, is increasingly proposed for verifying where footage was captured and for flagging manipulated or generated video. This letter shows that reading the mains frequency from a single short clip is confounded by the camera frame rate: at the rates in common use, a harmonic of that rate lands in the same narrow band as the true flicker, so the recovered grid is set by the frame rate, not the supply, and the estimator returns a confident value instead of abstaining. The underlying aliasing is known; its forensic consequence is not, and is the contribution here. On the 804 clips with a measurable in-band line the flicker feature does not separate genuine from generated video (area under the curve 0.509), while the recovered grid tracks the frame rate exactly for the generators, and a controlled synthesis confirms the mechanism against ground truth for genuine grids. Single-clip grid, and hence region, identification is therefore unreliable at the standard rates, and a comb-aware estimator that models the frame-rate harmonics restores the distinction between a genuine grid and a frame-rate comb under controlled conditions.***

## I. Introduction

THE mains supply imprints a faint periodic fluctuation, the electric network frequency (ENF), onto audio and, through the rolling-shutter mechanism, onto video [1]-[3]. This frequency drifts around its nominal value of 50 or 60 Hz in a manner characteristic of a regional grid, which has made it a valued forensic signature for timestamp verification [1] and region-of-recording identification [4]. An attractive extension is to read the nominal grid value from a single clip, with no reference database: a 50 Hz reading would place a recording on a 50 Hz grid and a 60 Hz reading on a 60 Hz grid, and, by extension, a generated clip carrying no consistent grid signature might be told apart from a genuine one.

The optical ENF is extracted from rolling-shutter video by exploiting the row-wise exposure [2], [3], with noise-resilient multi-harmonic estimation [5] and periodic zero-padding to correct the readout geometry [6], and has been used for region-of-recording identification [4], detecting the presence of an ENF trace [7], camera forensics [8] and video timestamping [9].

Manuscript submitted 14 October 2026. This work received no external funding. R. Wyton and B. Rehman are with Southampton Solent University, Southampton, U.K. (e-mail: 0wytor22@solent.ac.uk; bacha.rehman@solent.ac.uk). Code and data to reproduce the results is available at: https://github.com/RyanSolent95/enf-video-forensics

The rolling-shutter sampling geometry has been modelled in detail, including how the idle period between frames shifts and attenuates the observed ENF and how the frame rate governs the aliased component. For example, a 50 Hz supply aliases at 25 frames per second (fps), and a 60 Hz supply at 30 fps [10]. ENF has more recently been proposed as a physical cue for exposing manipulated video [11], [12], within a broader turn toward physical signals in forensic video analysis [13], and the field is surveyed in [14]. Prior work treats this frame-rate aliasing as a nuisance to be inverted for extraction and timestamping; its effect as a confound on single-clip grid identification, and on distinguishing genuine from generated video, is the question addressed here.

This letter shows that such single-clip grid identification is confounded by the frame rate: it is unreliable whenever the frame rate is degenerate with the mains frequency, that is when a harmonic of the frame rate falls in the 100 or 120 Hz flicker band. The aliasing itself is known from work that treats it as an extraction nuisance [10]; what is new here is establishing its forensic consequences, for region identification and for telling genuine from generated video, together with a mitigation. The contributions are as follows. First, the degeneracy that couples the frame rate to the recovered grid is set out (Section II). Second, on a survey of 1,072 clips it is shown that the flicker feature does not separate genuine from generated video and that the recovered grid is instead a deterministic function of the frame rate (Section III). Third, a controlled synthesis holds the true grid fixed while varying only the frame rate, confirming the mechanism against ground truth (Section III). Fourth, a comb-aware estimator that models the frame-rate harmonics is introduced as a mitigation (Section IV).

## II. The Frame-Rate Confound

A rolling-shutter sensor reads its rows sequentially; the exposure time of row $r$ in frame $f$ is given by:

$$t = \frac{f}{\text{fps}} + \frac{r\,T_{ro}}{H} \tag{1}$$

where $T_{ro}$ is the readout time and H the number of rows. The row rate $H/T_{ro}$ is far higher than the frame rate, so a light source modulated at twice the grid frequency, that is 100 Hz on a 50 Hz grid or 120 Hz on a 60 Hz grid, is resolved as spatial banding with energy at $2f_{grid}$ [3], [10]. Estimation therefore searches a narrow band, within ±2 Hz of 100 and 120 Hz, for the dominant spectral line.

The frame structure, however, also repeats at the frame rate. Any per-frame brightness variation, from automatic exposure or from the periodic structure of the video codec, is constant across a frame yet changes between frames, so on the reconstructed timeline it forms a comb with teeth at every integer multiple of the frame rate *fps*. When such a tooth falls inside the measurement band it is indistinguishable from a genuine flicker line:

$$k \cdot \text{fps} \approx 2f_{\text{grid}},\ k \in \mathbb{Z}^{+} \qquad (2)$$

where $k \in \mathbb{Z}^{+}$ (the positive integers) indexes the comb teeth and $f_{grid}$ is the grid frequency. Condition (2) holds when a comb tooth $k\cdot$fps falls within ±2 Hz of the flicker line at $2\cdot f_{grid}$, where it becomes indistinguishable from a genuine line.

At the frame rates in common use, condition (2) is met systematically rather than rarely (Table I). At 25 fps the fourth tooth is exactly 100 Hz, so any clip is read as a 50 Hz grid; at 24 and 30 fps the fifth and fourth teeth are 120 Hz, so any clip is read as 60 Hz. The recovered grid is therefore a function of the frame rate, not of the electricity supply, in line with the rolling-shutter sampling analysis of [10]. The row rate still resolves the true flicker, so the reading is reliable only when no tooth falls in the band, that is when the frame rate is non-degenerate with both 100 and 120 Hz.

TABLE I
FRAME-RATE DEGENERACY WITH THE ENF BAND

| Frame rate (fps) | In-band tooth | Apparent grid |
|---|---|---|
| 24 | 5 × 24 = 120 Hz | 60 Hz |
| 23.976 | 5 × 23.976 ≈ 120 Hz | 60 Hz |
| 25 | 4 × 25 = 100 Hz | 50 Hz |
| 29.97 | 4 × 29.97 ≈ 120 Hz | 60 Hz |
| 30 | 4 × 30 = 120 Hz | 60 Hz |
| 50 | 2 × 50 = 100 Hz | 50 Hz |
| 23, 27, 45 | none in band | unresolved |

## III. DATA AND RESULTS

### *A. Data*

A survey of 1,072 clips was assembled. The generated set comprises output from six text-to-video systems sampled from VBench-2.0 [15] (StepVideo, Kling, Sora, HunyuanVideo, Veo 3 and CogVideo); CogVideo renders at 16 fps, below the recovery envelope, and is excluded from the grid analysis. The genuine set spans both grid standards and a range of frame rates, drawn from a public rolling-shutter ENF dataset [16], the EV-ENFD dataset [17], studio recordings (RAVDESS [18]), MIT OpenCourseWare lectures (the MITFLD dataset [19]), and UK parliamentary proceedings from ParliamentLive.tv [20], together covering 50 Hz and 60 Hz supplies. Each clip is analysed over a segment of up to 120 s. For each clip the optical flicker was extracted with a second-harmonic estimator and two quantities were recorded: the prominence of the flicker line, as a genuine-versus-generated feature, and the recovered nominal grid. A focused analysis of 362 clips across thirteen sources supports the mechanism tests below.

### *B. The flicker feature does not separate real from generated video*

On the 804 clips that yielded a measurable in-band line, the flicker prominence does not separate genuine from generated video: the area under the receiver operating characteristic curve is 0.509 (95% confidence interval 0.469-0.549), indistinguishable from chance. Generated video carries an in-band line at real-comparable strength, so the presence of a strong 100 or 120 Hz line is not evidence of a genuine recording.

### *C. The recovered grid is fixed by the frame rate*

The recovered grid instead follows the frame rate exactly. Table II reports, for each generator, the proportion of clips read as 50 Hz rather than 60 Hz. Every clip from the 25 fps generator is read as 50 Hz, whereas the 24 and 30 fps generators are read as 50 Hz in at most 5% of clips (4 of 86, for Veo 3) and as 60 Hz otherwise (Fisher's exact test, $p < 10^{-40}$ between the 25 fps generator and each of the others). The effect is deterministic in the focused analysis: every clip captured at 24, 23.98, 29.97 or 30 fps returned 60 Hz, and every clip at 25 or 50 fps returned 50 Hz, across 346 clips without a single exception. The only clips that broke a one-grid-per-rate mapping were captured at non-degenerate rates, and there the genuine grid re-emerges: a real 60 Hz recording at 28 fps was read as 60 Hz, and the controlled synthesis below confirms the same for 50 Hz grids against ground truth.

TABLE II
RECOVERED GRID BY GENERATOR

| Generator | fps | Predicted | n | Read as 50 Hz | Read as 60 Hz |
|---|---|---|---|---|---|
| StepVideo | 25 | 50 | 100 | 100% | 0% |
| Kling | 24 | 60 | 100 | 0% | 100% |
| Sora | 30 | 60 | 100 | 1% | 99% |
| HunyuanVideo | 24 | 60 | 100 | 2% | 98% |
| Veo 3 | 24 | 60 | 86 | 5% | 95% |

CogVideo (16 fps) is excluded; its flicker aliases below the recovery envelope.

This is not a failure of the line estimator but a property of the signal. At a degenerate rate the in-band peak carries large absolute energy, with a floor-excess over empty frequencies of 2,000 to 4,400 for the 24 and 30 fps generators against an off-comb null of 36, yet it is statistically indistinguishable from the surrounding comb teeth: a leave-one-out test that refits the comb envelope without the in-band tooth and then scores it returns an excess of 1.05, below the 95th-percentile null of 3.25. The confident grid line is a frame-rate comb tooth, and the estimator is reading it correctly; it is simply not the grid. Fig. 1 contrasts the two cases: a genuine grid at a non-degenerate rate stands as a single line in the band, whereas a comb clip places a tooth there as one of a regular series.

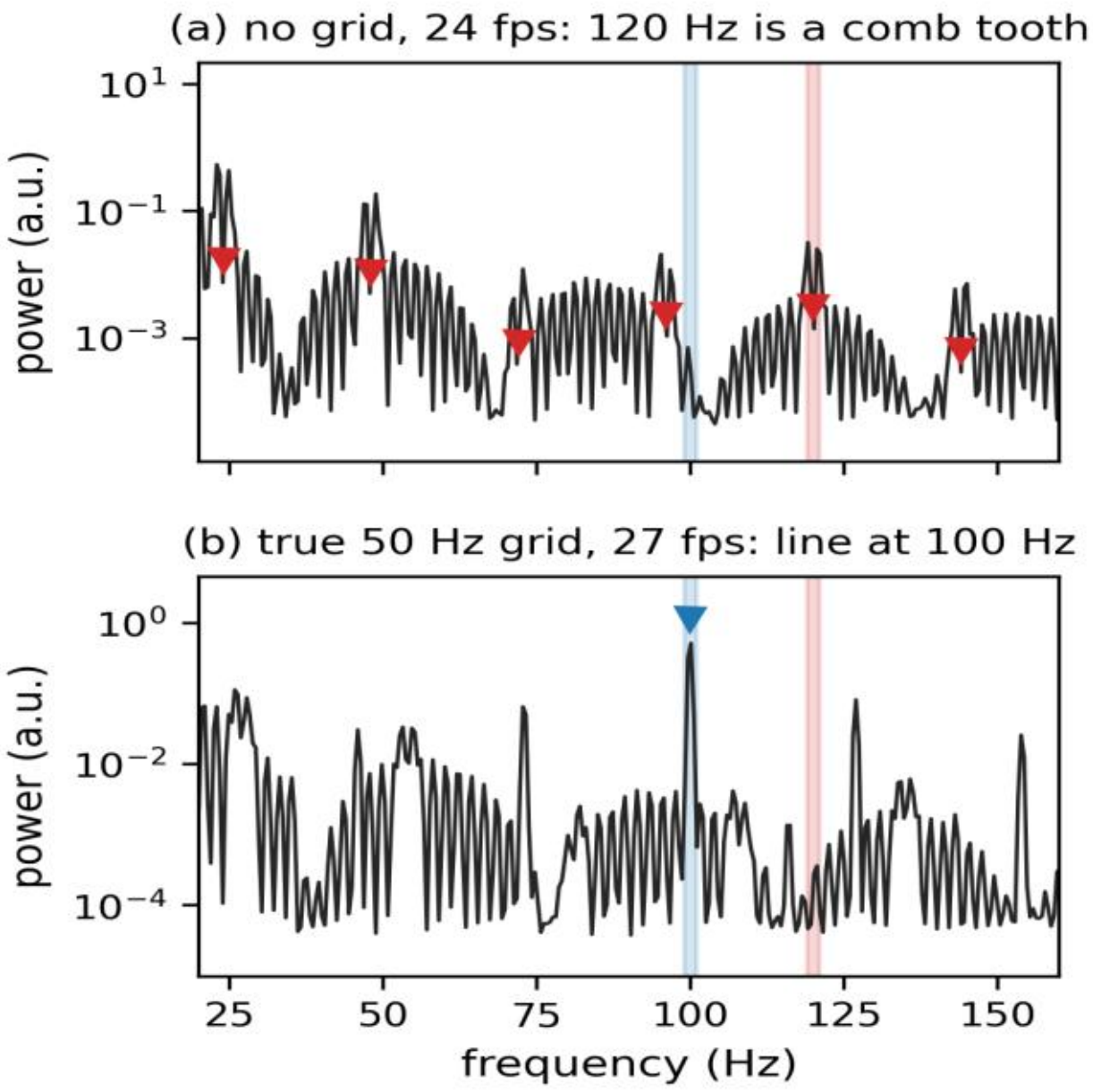

Fig. 1. Representative spectra (synthetic). (a) With no grid, at 24 fps the in-band peak at 120 Hz is one tooth of a regular frame-rate comb (triangles mark k × 24 Hz) and the 100 Hz band is empty. (b) A genuine 50 Hz grid at a non-degenerate 27 fps produces a single line at 100 Hz with no tooth beside it. The 100 Hz (blue) and 120 Hz (red) bands are shaded.

### *D. Controlled synthesis*

To confirm the mechanism against ground truth, clips were synthesised in which the true grid was held fixed while only the frame rate varied, across five random seeds per rate (Fig. 2). With no genuine grid present, the naive estimator returned a confident grid at every degenerate rate, matching the frame-rate prediction exactly (60 Hz at 24 and 30 fps, 50 Hz at 25 and 50 fps) and never abstaining. A genuine grid, by contrast, was read correctly at every frame rate on the majority vote. The confident wrong answer is therefore manufactured by the frame-rate comb in the absence of a grid, not by a weak but genuine signal.

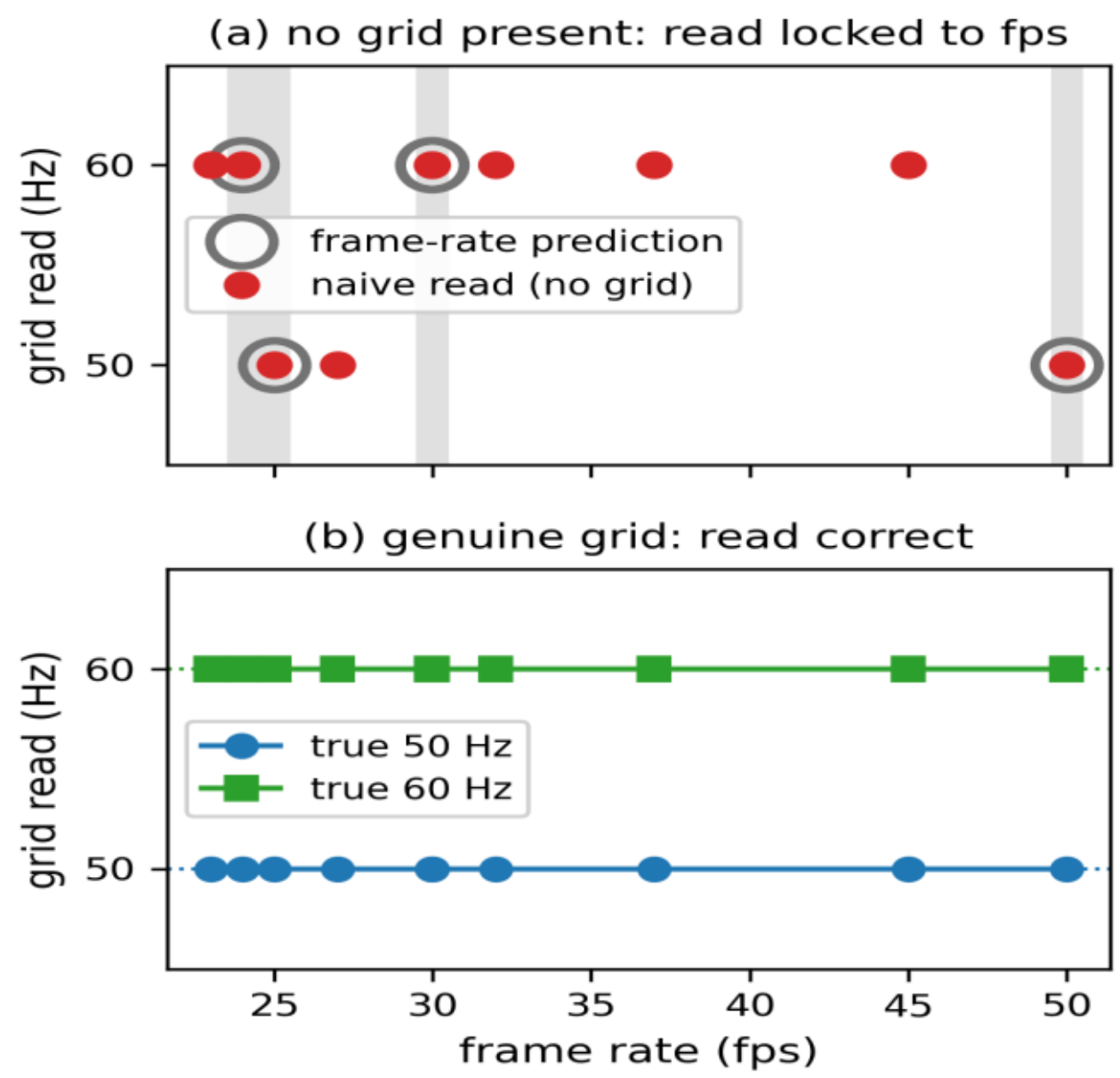

Fig. 2. Controlled synthesis, true grid fixed and only the frame rate varied. (a) With no grid present, the naive read is locked to the frame rate at every degenerate rate (shaded) and matches the frame-rate prediction, while wandering arbitrarily elsewhere. (b) A genuine grid reads correctly at every frame rate.

### *E. A temporal cross-check*

A further, weaker indicator is temporal. A frame-rate comb is locked to the camera clock, so its in-band tooth is strongly phase-coherent, whereas a genuine grid line wanders with the mains and is less coherent. Across the survey the generator teeth are, on average, more phase-coherent than genuine recordings captured at non-degenerate rates, which offers an estimator-independent hint rather than a decision on its own: a genuine recording captured at a degenerate rate has its own flicker coincide with a comb tooth and is then indistinguishable from one on this measure as well.

## IV. A Comb-Aware Mitigation

The confound can also be addressed in a learned optical detector used to flag generated video, a compact convolutional network trained by injecting a synthetic mains signature into flicker-free carriers and contrasting it with a set of manipulation classes. Three changes are used together. A wider spectrogram spans the second, fourth and sixth flicker harmonics rather than the second alone, so a genuine grid is corroborated across harmonics that a single frame-rate tooth does not populate. A relative prominence ratio is supplied as an explicit feature:

$$R = \frac{P_{100}}{P_{100} + P_{120}} \qquad (3)$$

where $P_{100}$ and $P_{120}$ are the energies in the 100 and 120 Hz bands. A genuine grid raises one band well clear of the other, whereas a comb leaves the opposite band at the noise floor. A dedicated comb class is added to training. Under controlled conditions the comb-aware detector rejects the frame-rate artefact that the baseline accepts as a genuine grid line: on 40 synthetic combs carrying no grid line, the deployed detector labels all 40 genuine while the comb-aware detector labels all 40 comb, with no genuine clip misclassified as a comb. This is a demonstrated mechanism rather than a solved problem: separating a faint genuine line from a comb on a single real clip still requires an external reference.

## V. Conclusion

Single-clip optical grid identification is governed by the camera frame rate rather than the mains supply, and it fails by returning a confident wrong answer instead of abstaining. The presence of a 100 or 120 Hz line is not evidence of authenticity, and a nominal grid read from a short clip is not assessable unless the frame rate is non-degenerate with both grid standards. In forensic practice, a single-clip grid reading should therefore be reported together with the frame rate, or withheld at degenerate rates. A comb-aware estimator that models the frame-rate harmonics restores the distinction under controlled conditions and, at non-degenerate rates, recovers the true grid. Closing the gap on degraded single clips, where a faint genuine line and a comb are not separable without a reference, is left to reference-matched training in future work.